\documentclass[]{aastex631}

\received{August 13, 2024}
\accepted{November 5, 2024}
\submitjournal{Astrophysical Journal Supplement}

\usepackage{graphicx}
\usepackage[caption=false]{subfig}
\usepackage{xcolor}

\newcommand{\ha}{H$\alpha$}

\newcommand{\kms}{km s$^{-1}$}

\shorttitle{Condor Observations of the M81 Group}
\shortauthors{Lanzetta et al.}

\begin{document}

\title{Introducing the Condor Array Telescope.  V.  Deep Broad- and Narrow-Band
Imaging Observations of the M81 Group}

\author{Kenneth M.\ Lanzetta}
\affiliation{Department of Physics and Astronomy,
Stony Brook University,
Stony Brook, NY 11794-3800, USA}

\author{Stefan Gromoll}
\affiliation{Amazon Web Services,
410 Terry Ave.\ N,
Seattle, WA 98109, USA}

\author{Michael M.\ Shara}
\affiliation{Department of Astrophysics,
American Museum of Natural History,
Central Park West at 79th St.,
New York, NY 10024-5192, USA}

\author{David Valls-Gabaud}
\affiliation{Observatoire de Paris,
LERMA, CNRS,
61 Avenue de l'Observatoire,
75014 Paris, FRANCE}

\author{Frederick M.\ Walter}
\affiliation{Department of Physics and Astronomy,
Stony Brook University,
Stony Brook, NY 11794-3800, USA}

\author{John K. Webb}
\affiliation{Institute of Astronomy,
University of Cambridge,
Madingley Road,
Cambridge CB3 0HA, UNITED KINGDOM}

\begin{abstract}
We used the Condor Array Telescope to obtain deep imaging observations through
the luminance broad-band and He~II 468.6 nm, [O~III] 500.7 nm, He~I 587.5 nm,
\ha, [N~II] 658.4 nm, and [S~II] 671.6 nm narrow-band filters of an extended
region comprising 13 ``Condor fields'' spanning $\approx 8 \times 8$ deg$^2$ on
the sky centered near M81 and M82.  Here we describe the acquisition and
processing of these observations, which together constitute unique very deep
imaging observations of a large portion of the M81 Group through a complement
of broad- and narrow-band filters.  The images are characterized by an
intricate web of faint, diffuse, continuum produced by starlight scattered from
Galactic cirrus, and all prominent cirrus features identified in the broad-band
image can also be identified in the narrow-band images.  We subtracted the
luminance image from the narrow-band images to leave more or less only line
emission in the difference images, and we masked regions of the resulting
images around stars at an isophotal limit.  The difference images exhibit
extensive extended structures of ionized gas in the direction of the M81 Group,
from known galaxies of the M81 Group, clouds of gas, filamentary structures,
and apparent or possible bubbles or shells.  Specifically, the difference
images show a remarkable filament known as the ``Ursa Major Arc;'' a remarkable
network of criss-crossed filaments between M81 and NGC 2976, some of which
intersect and overlap the Ursa Major Arc; and details of a ``giant shell of
ionized gas.''
\end{abstract}


\section{Introduction}

The M81 Group consists of the galaxies M81, M82, NGC 3077, NGC 2976, and more
than 30 other galaxies \citep{kar2005} that together stretch for more than 10
deg on the sky.  The group is located just 3.6 Mpc away \citep{fre1994}, making
it one of the nearest galaxy groups beyond the Local Group.  The group is
notable in that its most prominent members M81, M82, and NGC 3077 are strongly
interacting, and this interaction has caused gas and stars to be stripped from
the galaxies into a complex network of gaseous and stellar streams and has
driven strong star-formation activity in the centers of M82 and NGC 3007.  In
fact, M82 is the prototypical and by far nearest example of a ``starburst''
galaxy with a high-velocity ``superwind'' that emanates from its center.

Analysis and interpretation of all deep imaging observations of the M81 Group
are made complicated by two properties of the group:  (1) The group exhibits a
very low recession velocity, with M81 itself exhibiting a negative heliocentric
recession velocity of $-36$ \kms\ \citep{kou2020}.  This makes it difficult to
distinguish features of Galactic origin from features at the distance of the
group, even with spectroscopic observations in hand.  And (2) the group is
located at a Galactic latitude of $b \approx 40$ deg in a direction of
significant Galactic cirrus.  Starlight scattered from the cirrus dominates the
diffuse light seen in {\em both} broad- and narrow-band images, thereby
muddying the view of foreground and background objects alike.

Over a period stretching from December 2021 through February 2024, we used the
Condor Array Telescope \citep{lan2023} to obtain deep imaging observations
through the luminance broad-band and He~II 468.6 nm, [O~III] 500.7 nm, He~I
587.5 nm, H$\alpha$, [N~II] 658.4 nm, and [S~II] 671.6 nm narrow-band filters
of an extended region comprising 13 ``Condor fields''\footnote{``Condor
fields'' are a set of fields with field centers that tile the entire sky with
the Condor field of view, allowing for overlap.} spanning $\approx 8 \times 8$
deg$^2$ on the sky centered near M81 and M82.  We obtained all broad-band
observations at a cadence of one minute.  Our motivation was severalfold:
\begin{itemize}

\item to characterize the Galactic cirrus in the direction of the M81 group to
very sensitive levels;

\item to search for ionized gas in the direction of the M81 Group, perhaps
associated with the group or the cosmic web or perhaps associated with the
Galaxy;

\item to experiment with possibilities of combining deep imaging observations
through both broad- and narrow-band filters to determine and subtract continuum
from narrow-band images in regions of significant Galactic cirrus;

\item to exploit the diagnostic capabilities of narrow-band images obtained
through the [O~III], \ha, [N~II], and [S~II] filters to determine or constrain
physical properties of ionized gas;

 \item to use the broad-band observations obtained at a one-minute cadence to
search for short-period planets transiting white dwarfs;

\item to use the broad-band observations summed at a daily cadence to search
for intergalactic novae in the outskirts of and between and around M81, M82,
NGC 3077, NGC 2976, and other galaxies of the group;

\item to use the broad- and narrow-band observations to search for
intergalactic planetary nebulae in the outskirts of and between and around M81,
M82, NGC 3077, NGC 2976, and other galaxies of the group;

\item and to continue to establish and demonstrate the low-surface-brightness
sensitivity of Condor through both broad- and narrow-band filters.

\end{itemize}

Here we describe the acquisition and processing of these observations, which
we believe together constitute the deepest wide-field imaging observations of
the M81 Group ever obtained through a complement of broad- and narrow-band
filters.  We use these observations to examine ionized gas in the direction of
the M81 Group in \citet{lan2024b}, and we will present additional results of
the analysis and interpretation of these observations elsewhere, addressing
various of the topics noted above.

\section{Observations}

Condor is an ``array telescope'' that consists of six apochromatic refracting
telescopes of objective diameter 180 mm, each equipped with a large-format
($9576 \times 6388$ pixels), very low-read-noise ($\approx 1.2$ e$^-$), very
rapid-read-time ($< 1$ s) CMOS camera.  Details of the motivation,
configuration, and performance of the telescope are described by
\citet{lan2023}.

Over a period stretching from December 2021 through February 2024, we used
Condor to obtain deep imaging observations through the luminance, He~II 468.6
nm, [O~III] 500.7 nm, He~I 587.5 nm, H$\alpha$ 656.3 nm, [N~II] 658.4 nm, and
[S~II] 671.6 nm narrow-band filters of an extended region comprising 13
``Condor fields'' spanning $\approx 8 \times 8$ deg$^2$ on the sky centered
near M81 and M82.  All exposures through the luminance filter were obtained
with individual exposure times of 60 s, and all exposures through the
narrow-band filters were obtained with individual exposure times of 600 s.  The
telescope was dithered by a random offset of $\approx 15$ arcmin between each
exposure.

Describing the resources devoted to a particular observation using an array
telescope like Condor can be difficult in that the individual telescopes that
comprise the array might be used in various filter configurations, with one or
more telescopes allocated to a particular filter for a particular observation.
(Further, various of the individual telescope might fail or be otherwise
unavailable from time to time, temporarily taking them out of the mix.)  To
address this issue, we introduce here the concept of the ``reach'' of an
observation obtained by an array (or ordinary) telescope, which we define as
the product of the total objective area and the total exposure time devoted to
the observation.  In its current configuration, Condor consists of six
individual telescopes, each of objective diameter $180$ mm or objective area
$0.0254$ m$^2$.  Hence a one-second exposure with one telescope of the array
yields a reach of $0.0254$ m$^2$ s, and a one-second exposure with the entire
array (i.e.\ with all six telescopes of the array) yields a reach of $6 \times
0.0254 = 0.153$ m$^2$ s.  Because Condor is used with variety of filter
configurations, with between one and six telescopes allocated to a particular
filter for a particular observation, and because it is of interest to compare
the capabilities of Condor with those of other telescopes, it is appropriate to
employ the concept of the reach to concisely describe observations obtained
using array telescopes.

Details of the observations are presented in Table 1, which for each pointing
lists the Condor field identifier, the International Celestial Reference System
(ICRS) J2000 coordinates of the field center, the filter, the start and end
dates of the observations, and the reach as described above.

\begin{table}[ht]
\centering
\hspace{-0.80in}
\begin{tabular}{p{1.00in}cccccr}
\multicolumn{7}{c}{{\bf Table 1:}  Details of Observations} \\
\hline
\hline
\multicolumn{1}{c}{} & \multicolumn{2}{c}{J2000} &
\multicolumn{1}{c}{} & \multicolumn{1}{c}{Start} &
\multicolumn{1}{c}{End} & \multicolumn{1}{c}{Reach} \\
\cline{2-3}
\multicolumn{1}{c}{Condor Field} & \multicolumn{1}{c}{R.A.} &
\multicolumn{1}{c}{Dec} & \multicolumn{1}{c}{Filter} &
\multicolumn{1}{c}{Date} & \multicolumn{1}{c}{Date} &
\multicolumn{1}{c}{(m$^2$ s)} \\
\hline
06917 \dotfill & 09:25:42.86 & $+$70:54:32.76 &
      luminance           & 2021-11-30 & 2022-03-20 & 7,702.8 \\
\multicolumn{1}{c}{} & & & He II 468.6 nm      & 2021-12-04 & 2022-03-16 & 519.1 \\
\multicolumn{1}{c}{} & & & [O III] 500.7 nm    & 2021-12-04 & 2024-02-02 & 664.2 \\
\multicolumn{1}{c}{} & & & He I 587.5 nm       & 2021-12-04 & 2022-03-16 & 351.2 \\
\multicolumn{1}{c}{} & & & H-alpha 656.3 nm    & 2021-12-04 & 2024-02-02 & 790.1 \\
\multicolumn{1}{c}{} & & & [N II] 658.4 nm     & 2021-12-04 & 2022-03-16 & 412.2 \\
\multicolumn{1}{c}{} & & & [S II] 671.6 nm     & 2021-11-30 & 2024-02-02 & 224.4 \\
06730 \dotfill & 09:31:45.89 & $+$66:49:05.52 &
      luminance           & 2021-12-01 & 2022-03-25 & 7,737.9 \\
\multicolumn{1}{c}{} & & & He II 468.6 nm      & 2021-12-04 & 2022-03-16 & 458.0 \\
\multicolumn{1}{c}{} & & & [O III] 500.7 nm    & 2021-12-04 & 2024-01-07 & 618.4 \\
\multicolumn{1}{c}{} & & & He I 587.5 nm       & 2021-12-04 & 2022-03-16 & 412.2 \\
\multicolumn{1}{c}{} & & & H-alpha 656.3 nm    & 2021-12-04 & 2024-02-02 & 757.0 \\
\multicolumn{1}{c}{} & & & [N II] 658.4 nm     & 2021-12-04 & 2023-04-28 & 488.6 \\
\multicolumn{1}{c}{} & & & [S II] 671.6 nm     & 2021-12-04 & 2024-01-07 & 313.0 \\
06859 \dotfill & 09:36:00.00 & $+$69:32:43.80 &
      luminance           & 2021-11-30 & 2022-03-25 & 7,905.8 \\
\multicolumn{1}{c}{} & & & He II 468.6 nm      & 2021-12-04 & 2022-03-16 & 503.8 \\
\multicolumn{1}{c}{} & & & [O III] 500.7 nm    & 2021-12-04 & 2024-02-02 & 694.7 \\
\multicolumn{1}{c}{} & & & He I 587.5 nm       & 2021-12-04 & 2022-03-16 & 488.6 \\
\multicolumn{1}{c}{} & & & H-alpha 656.3 nm    & 2021-12-04 & 2024-02-01 & 806.7 \\
\multicolumn{1}{c}{} & & & [N II] 658.4 nm     & 2021-12-04 & 2023-04-28 & 519.1 \\
\multicolumn{1}{c}{} & & & [S II] 671.6 nm     & 2021-12-04 & 2024-02-01 & 267.2 \\
06972 \dotfill & 09:41:32.30 & $+$72:16:21.72 &
      luminance           & 2021-12-01 & 2022-03-25 & 6,562.2 \\
\multicolumn{1}{c}{} & & & He II 468.6 nm      & 2021-12-04 & 2022-03-16 & 412.2 \\
\multicolumn{1}{c}{} & & & [O III] 500.7 nm    & 2021-12-04 & 2024-01-07 & 572.6 \\
\multicolumn{1}{c}{} & & & He I 587.5 nm       & 2021-12-04 & 2022-03-16 & 366.4 \\
\multicolumn{1}{c}{} & & & H-alpha 656.3 nm    & 2021-12-04 & 2024-01-07 & 759.6 \\
\multicolumn{1}{c}{} & & & [N II] 658.4 nm     & 2021-12-04 & 2023-04-28 & 442.8 \\
\multicolumn{1}{c}{} & & & [S II] 671.6 nm     & 2021-12-04 & 2024-01-07 & 251.9 \\
06797 \dotfill & 09:45:00.00 & $+$68:10:54.48 &
      luminance           & 2021-11-30 & 2022-03-25 & 10,582.3 \\
\multicolumn{1}{c}{} & & & He II 468.6 nm      & 2021-12-04 & 2022-03-16 & 320.6 \\
\multicolumn{1}{c}{} & & & [O III] 500.7 nm    & 2021-12-04 & 2024-02-01 & 450.4 \\
\multicolumn{1}{c}{} & & & He I 587.5 nm       & 2021-12-04 & 2022-03-16 & 366.4 \\
\multicolumn{1}{c}{} & & & H-alpha 656.3 nm    & 2021-12-04 & 2024-02-01 & 713.8 \\
\multicolumn{1}{c}{} & & & [N II] 658.4 nm     & 2021-12-04 & 2023-04-28 & 366.4 \\
\multicolumn{1}{c}{} & & & [S II] 671.6 nm     & 2021-12-04 & 2024-02-01 & 221.4 \\
06918 \dotfill & 09:51:25.70 & $+$70:54:32.76 &
      luminance           & 2021-12-01 & 2022-03-25 & 8,825.0 \\
\multicolumn{1}{c}{} & & & He II 468.6 nm      & 2021-12-04 & 2022-03-16 & 335.9 \\
\multicolumn{1}{c}{} & & & [O III] 500.7 nm    & 2021-12-04 & 2024-02-01 & 482.2 \\
\multicolumn{1}{c}{} & & & He I 587.5 nm       & 2021-12-04 & 2022-03-16 & 366.4 \\
\multicolumn{1}{c}{} & & & H-alpha 656.3 nm    & 2021-12-04 & 2024-02-01 & 698.5 \\
\multicolumn{1}{c}{} & & & [N II] 658.4 nm     & 2021-12-04 & 2023-04-28 & 335.9 \\
\multicolumn{1}{c}{} & & & [S II] 671.6 nm     & 2021-12-04 & 2024-02-01 & 206.1 \\
06731 \dotfill & 09:52:56.47 & $+$66:49:05.52 &
      luminance           & 2021-11-30 & 2022-03-25 & 9,683.0 \\
\multicolumn{1}{c}{} & & & He II 468.6 nm      & 2021-12-04 & 2022-03-16 & 335.9 \\
\multicolumn{1}{c}{} & & & [O III] 500.7 nm    & 2021-12-04 & 2024-02-01 & 465.7 \\
\multicolumn{1}{c}{} & & & He I 587.5 nm       & 2021-12-04 & 2022-03-16 & 351.2 \\
\multicolumn{1}{c}{} & & & H-alpha 656.3 nm    & 2021-12-04 & 2024-02-01 & 606.9 \\
\multicolumn{1}{c}{} & & & [N II] 658.4 nm     & 2021-12-04 & 2023-04-28 & 351.2 \\
\multicolumn{1}{c}{} & & & [S II] 671.6 nm     & 2021-12-04 & 2024-02-01 & 161.6 \\
\hline
\end{tabular}
\end{table}

\begin{table}[ht]
\centering
\hspace{-0.80in}
\begin{tabular}{p{1.00in}cccccr}
\multicolumn{7}{c}{{\bf Table 1} (continued)} \\
\hline
\hline
\multicolumn{1}{c}{} & \multicolumn{2}{c}{J2000} &
\multicolumn{1}{c}{} & \multicolumn{1}{c}{Start} &
\multicolumn{1}{c}{End} & \multicolumn{1}{c}{Reach} \\
\cline{2-3}
\multicolumn{1}{c}{Condor Field} & \multicolumn{1}{c}{R.A.} &
\multicolumn{1}{c}{Dec} & \multicolumn{1}{c}{Filter} &
\multicolumn{1}{c}{Date} & \multicolumn{1}{c}{Date} &
\multicolumn{1}{c}{(m$^2$ s)} \\
\hline
06860 \dotfill & 10:00:00.00 & $+$69:32:43.80 &
      luminance           & 2021-12-01 & 2023-05-25 & 15,799.5 \\
\multicolumn{1}{c}{} & & & He II 468.6 nm      & 2021-12-04 & 2023-02-27 & 1,710.0 \\
\multicolumn{1}{c}{} & & & [O III] 500.7 nm    & 2021-12-04 & 2024-02-02 & 3,137.6 \\
\multicolumn{1}{c}{} & & & He I 587.5 nm       & 2021-12-04 & 2023-02-27 & 1,786.4 \\
\multicolumn{1}{c}{} & & & H-alpha 656.3 nm    & 2021-12-04 & 2024-02-04 & 622.2 \\
\multicolumn{1}{c}{} & & & [N II] 658.4 nm     & 2021-12-04 & 2023-04-28 & 335.9 \\
\multicolumn{1}{c}{} & & & [S II] 671.6 nm     & 2021-12-04 & 2024-02-04 & 160.3 \\
06798 \dotfill & 10:07:30.00 & $+$68:10:54.48 &
      luminance           & 2021-11-30 & 2022-03-25 & 8,246.3 \\
\multicolumn{1}{c}{} & & & He II 468.6 nm      & 2021-12-04 & 2022-03-16 & 229.0 \\
\multicolumn{1}{c}{} & & & [O III] 500.7 nm    & 2021-12-04 & 2024-01-01 & 271.0 \\
\multicolumn{1}{c}{} & & & He I 587.5 nm       & 2021-12-04 & 2022-03-16 & 229.0 \\
\multicolumn{1}{c}{} & & & H-alpha 656.3 nm    & 2021-12-04 & 2024-01-01 & 458.0 \\
\multicolumn{1}{c}{} & & & [N II] 658.4 nm     & 2021-12-04 & 2023-04-28 & 244.3 \\
\multicolumn{1}{c}{} & & & [S II] 671.6 nm     & 2021-12-04 & 2024-01-01 & 87.8 \\
06973 \dotfill & 10:09:13.85 & $+$72:16:21.72 &
      luminance           & 2021-12-01 & 2022-03-25 & 7,380.6 \\
\multicolumn{1}{c}{} & & & He II 468.6 nm      & 2021-12-04 & 2022-03-16 & 244.3 \\
\multicolumn{1}{c}{} & & & [O III] 500.7 nm    & 2021-12-04 & 2024-01-01 & 272.3 \\
\multicolumn{1}{c}{} & & & He I 587.5 nm       & 2021-12-04 & 2022-03-16 & 259.6 \\
\multicolumn{1}{c}{} & & & H-alpha 656.3 nm    & 2021-12-04 & 2024-01-01 & 459.3 \\
\multicolumn{1}{c}{} & & & [N II] 658.4 nm     & 2021-12-04 & 2023-04-28 & 290.1 \\
\multicolumn{1}{c}{} & & & [S II] 671.6 nm     & 2021-12-04 & 2024-01-01 & 87.8 \\
06732 \dotfill & 10:14:07.06 & $+$66:49:05.52 &
      luminance           & 2021-11-30 & 2022-03-25 & 7,473.7 \\
\multicolumn{1}{c}{} & & & He II 468.6 nm      & 2021-12-04 & 2022-03-16 & 213.8 \\
\multicolumn{1}{c}{} & & & [O III] 500.7 nm    & 2021-12-04 & 2024-01-01 & 255.7 \\
\multicolumn{1}{c}{} & & & He I 587.5 nm       & 2021-12-04 & 2022-03-16 & 229.0 \\
\multicolumn{1}{c}{} & & & H-alpha 656.3 nm    & 2021-12-04 & 2024-01-01 & 442.8 \\
\multicolumn{1}{c}{} & & & [N II] 658.4 nm     & 2021-12-04 & 2023-04-28 & 229.0 \\
\multicolumn{1}{c}{} & & & [S II] 671.6 nm     & 2021-12-04 & 2024-01-01 & 103.1 \\
06919 \dotfill & 10:17:08.57 & $+$70:54:32.76 &
      luminance           & 2021-12-03 & 2022-03-25 & 3,264.3 \\
\multicolumn{1}{c}{} & & & He II 468.6 nm      & 2021-12-04 & 2022-03-16 & 183.2 \\
\multicolumn{1}{c}{} & & & [O III] 500.7 nm    & 2021-12-04 & 2024-01-01 & 258.3 \\
\multicolumn{1}{c}{} & & & He I 587.5 nm       & 2021-12-04 & 2022-03-16 & 106.9 \\
\multicolumn{1}{c}{} & & & H-alpha 656.3 nm    & 2021-12-04 & 2024-01-01 & 398.2 \\
\multicolumn{1}{c}{} & & & [N II] 658.4 nm     & 2021-12-04 & 2023-04-28 & 198.5 \\
\multicolumn{1}{c}{} & & & [S II] 671.6 nm     & 2021-12-04 & 2024-01-01 & 87.8 \\
06861 \dotfill & 10:24:00.00 & $+$69:32:43.80 &
      luminance           & 2021-12-02 & 2022-03-25 & 6,078.2 \\
\multicolumn{1}{c}{} & & & He II 468.6 nm      & 2021-12-04 & 2022-03-16 & 198.5 \\
\multicolumn{1}{c}{} & & & [O III] 500.7 nm    & 2021-12-04 & 2024-01-01 & 286.3 \\
\multicolumn{1}{c}{} & & & He I 587.5 nm       & 2021-12-04 & 2022-03-16 & 183.2 \\
\multicolumn{1}{c}{} & & & H-alpha 656.3 nm    & 2021-12-04 & 2024-01-01 & 412.2 \\
\multicolumn{1}{c}{} & & & [N II] 658.4 nm     & 2021-12-04 & 2023-04-28 & 213.8 \\
\multicolumn{1}{c}{} & & & [S II] 671.6 nm     & 2021-12-04 & 2024-01-01 & 87.8 \\
\hline
\end{tabular}
\end{table}

For comparison, the reach of narrow-band imaging observations of the galaxies
M81 and M82 recently obtained by \citet{lok2022} using the Dragonfly Spectral
Line pathfinder are $6.9 \times 10^3$ m$^2$ s through the \ha\ filter and $2.7
\times 10^3$ m$^2$ s through the [N~II] 658.4 nm filter, over a field of view
spanning $2 \times 3$ deg$^2$.

\section{Data Processing}

We processed the observations using the Condor data pipeline, which is
described by \citet{lan2023} and \citet{lan2023a}.  Briefly, the data pipeline
performs bias subtraction; field flattening and background subtraction;
astrometric calibration; identification and masking of cosmic ray events,
satellite trails, and pixels that exhibit significant random telegraph noise;
photometric calibration; drizzling onto a common coordinate grid; and
coaddition of the individual exposures.  After this processing, we combined the
various groups of coadded images into mosaic images, yielding a total of seven
mosaic images (one each obtained through the luminance and six narrow-band
filters).

The resulting mosaic images obtained through the luminance and \ha\ filters are
shown in Figures 1 and 2, as examples of the observations.  All mosaic images
(through the luminance, He~II, [O~III], He~I, \ha, [N~II], and [S~II] filters)
as well as all individual coadded images obtained through all filters are
available for download on the Condor web site, as described in \S\ 8 below.
Each mosaic image contains $\approx 34,000 \times 34,000$ pixels (i.e.\ around
a billion pixels) at a pixel scale of $0.85$ arcsec pix$^{-1}$.  Yet our
primary interest here is on angular scales that are much larger than the angle
subtended by a single pixel.  Accordingly, the mosaic images are displayed here
block averaged by $32 \times 32$ pixels, which yields an effective image size
of $\approx 1000 \times 1000$ pixels at an effective pixel scale of $27.2$
arcsec pix$^{-1}$.

We note several aspects of the mosaic images shown in Figures 1 and 2 and the
other individual and mosaic coadded images as follows:
\begin{enumerate}

\item Because the reach varies between the individual coadded images (as
presented in Table 1), the sensitivities of the mosaic images vary across the
images.  These variations are completely tracked by the uncertainty images
associated with the individual coadded images that are produced by the data
pipeline.

\item Due to technical difficulties with a filter wheel, there is a dearth of
exposures obtained through the [S~II] filter over the region covering roughly
the left half of the mosaic image, and the exposures that were obtained over
this region are of substandard quality.  Although we present these data for the
sake of completeness, we note that this region of the [S~II] mosaic image
should be excluded from any quantitative measurements.

\item A few very faint satellite trails survived the data processing,
particularly in regions of images containing fewer than average exposures.

\item The images obtained through the He II and He I filters do not appear to
show significant line emission.

\end{enumerate}

\begin{figure}[ht!]
\centering
\subfloat{
  \includegraphics[width=0.58\linewidth, angle=0]{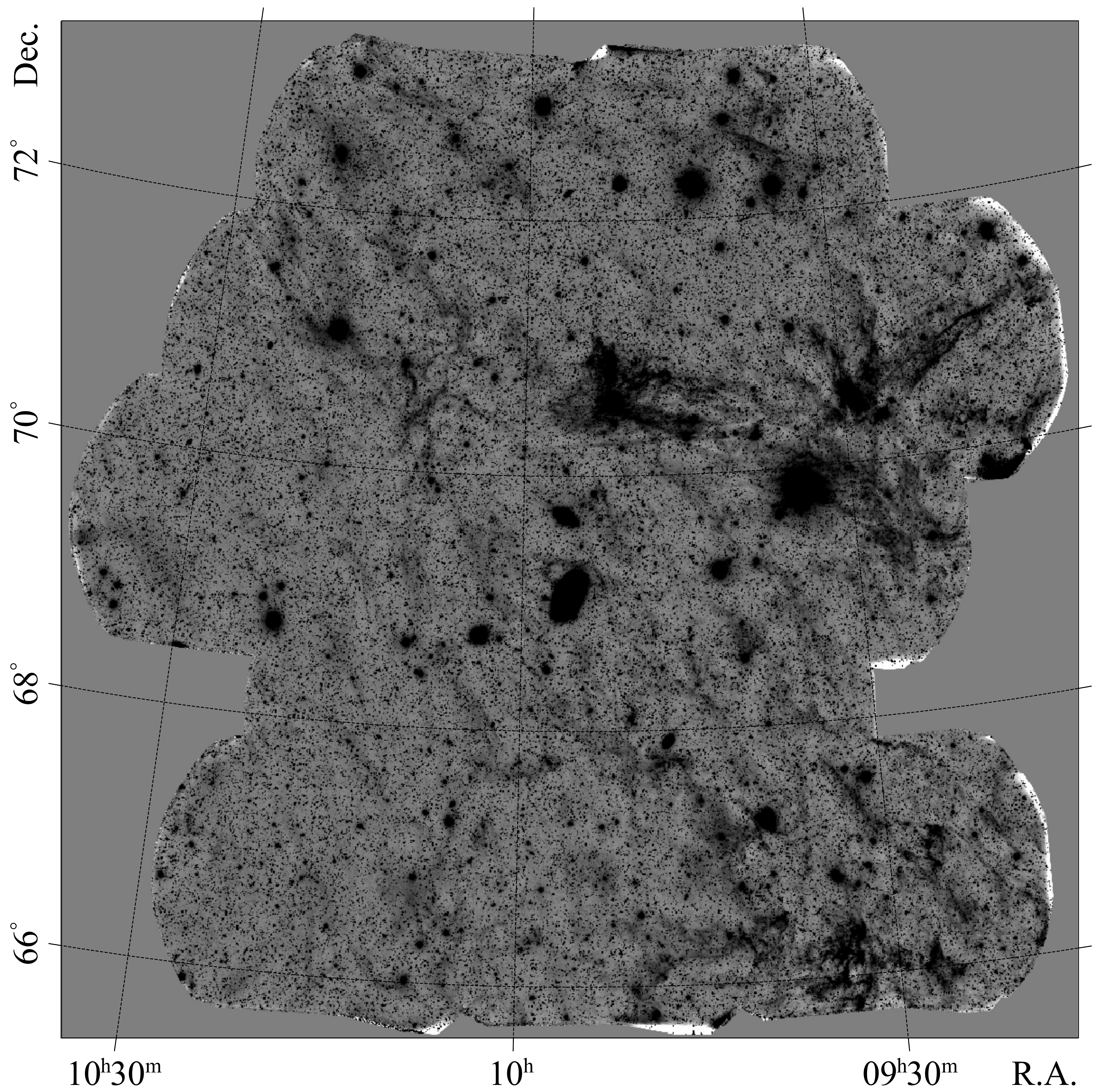}
}
\caption{Mosaic image of M81 Group obtained through luminance filter.  Image
is displayed block averaged by $32 \times 32$ pixels.}
\end{figure}

\begin{figure}[ht!]
\centering
\subfloat{
  \includegraphics[width=0.58\linewidth, angle=0]{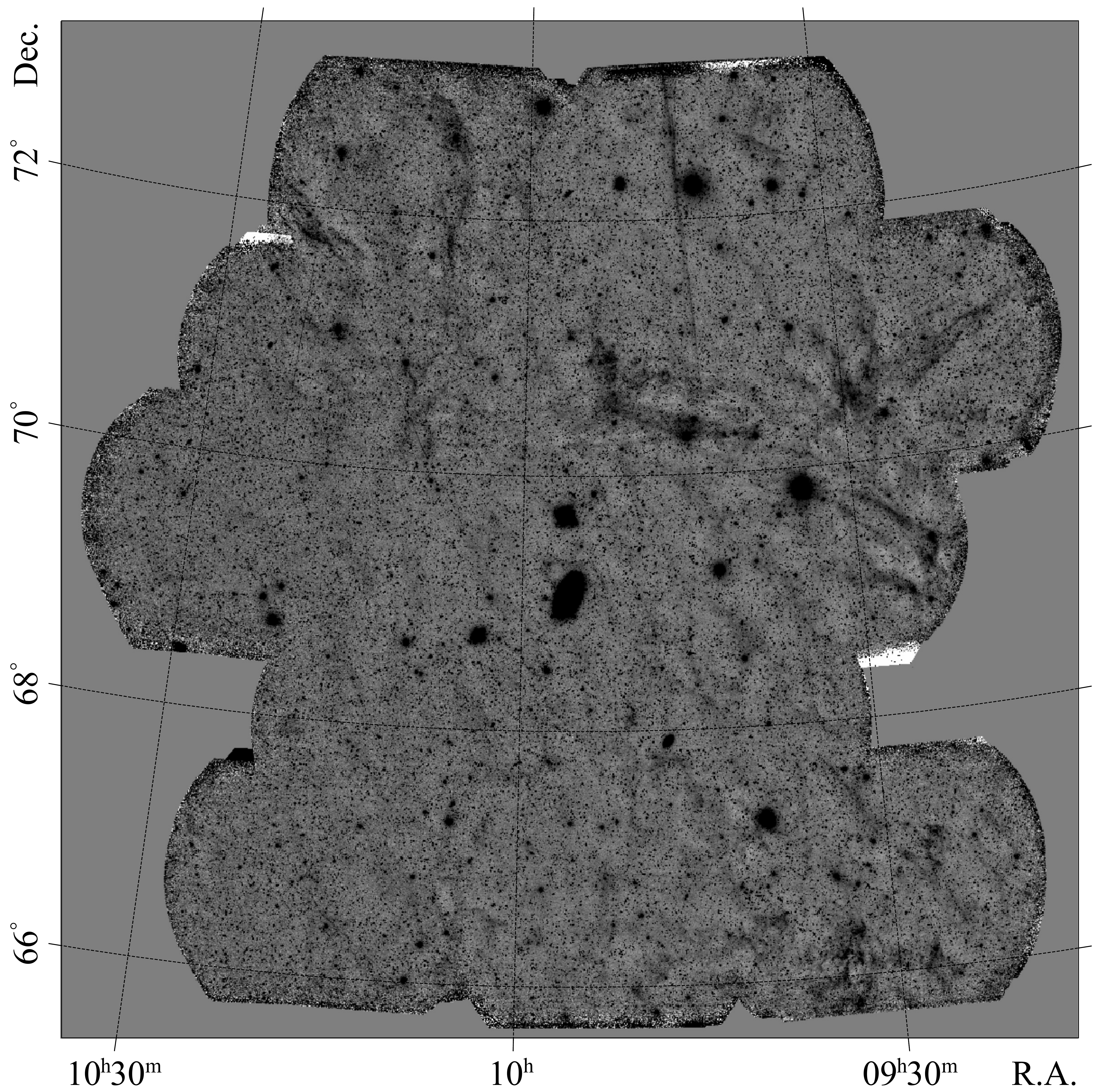}
}
\caption{Mosaic image of M81 Group obtained through \ha\ filter.  Image is
displayed block averaged by $32 \times 32$ pixels.}
\end{figure}

The images shown in Figures 1 and 2 and the other mosaic and individual coadded
images together constitute the deepest wide-field imaging observations of the
M81 Group yet obtained through a complement of broad- and narrow-band filters.
In terms of depth, angular resolution, and field of view, the luminance image
provides an unprecedented view of the Galactic cirrus toward the M81 Group,
while the narrow-band images provide an unprecedented view of ionized gas
toward the M81 Group.  The maximum $3 \sigma$ surface-brightness sensitivity of
the luminance image is $1.8 \times 10^{-31}$ erg s$^{-1}$ cm$^{-2}$ Hz$^{-1}$
arcsec$^{-2}$ over $10 \times 10$ arcsec$^2$ regions and $6.9 \times 10^{-32}$
erg s$^{-1}$ cm$^{-2}$ Hz$^{-1}$ arcsec$^{-2}$ over $32 \times 32$ pix$^2$ (or
$27.2 \times 27.2$ arcsec$^2$) regions, and the maximum $3 \sigma$
surface-brightness sensitivities of the narrow-band images regions are
presented in Table 2 (recognizing that the sensitivities of the images vary
across the images).

\begin{table}[ht]
\centering
\hspace{-0.80in}
\begin{tabular}{p{1.75in}ccc}
\multicolumn{4}{c}{{\bf Table 2:}  Maximum Surface Brightness Sensitivities} \\
\hline
\hline
\multicolumn{1}{c}{} & \multicolumn{3}{c}{$3 \sigma$ sensitivity} \\
\multicolumn{1}{c}{} & \multicolumn{3}{c}{($10^{-18}$ erg s$^{-1}$ cm$^{-2}$
arcsec$^{-2}$)} \\
\cline{2-4}
\multicolumn{1}{c}{Filter} & $10 \times 10$ arcsec$^2$ & & $32 \times 32$ pix$^2$ \\
\hline
He II 468.6 nm \dotfill     & 3.0 & & 1.1 \\
{[O III]} 500.7 nm \dotfill & 6.6 & & 2.4 \\
He I 587.5 nm \dotfill      & 2.1 & & 0.78 \\
\ha\ \dotfill               & 3.9 & & 1.4 \\
{[N II]} 658.4 nm \dotfill  & 9.6 & & 3.6  \\
{[S II]} 671.6 nm \dotfill  & 11  & & 3.9  \\
\hline
\end{tabular}
\end{table}

\section{Photometric Calibration}

The Condor data pipeline uses measurements of continuum sources (i.e.\ stars)
to determine photometric calibrations of all broad- and narrow-band images
in terms of monochromatic energy flux densities \citep{lan2023, lan2023a}.  For
measuring emission-line sources, it is more appropriate to express the
calibration of the narrow-band images in terms of bolometric energy fluxes,
integrated across the bandpasses of the filters.  The [O~III], \ha, [N~II], and
[S~II] have bandpasses of ${\rm FWHM} \approx 3$ nm width, while the He~II and
He~I filters have bandpasses of ${\rm FWHM} \approx 4$ nm width, and the
response functions of all six narrow-band filters are characterized by very
steep edges \citep{lan2023}.  Accordingly, we determined the photometric
calibrations of the narrow-band images in terms of bolometric energy fluxes
simply by multiplying the monochromatic energy flux densities by the nominal
bandpasses of the filters.  (The corresponding velocity widths of the filters
are typically $\approx 1500$ \kms\ for the narrower filters and $\approx 2000$
\kms\ for the broader filters, i.e.\ much larger than the recession velocity of
the M81 Group.)

To assess the accuracy of this procedure, we compared measurements of the
bolometric energy fluxes of the \ha\ plus [N~II] 658.4 nm emission lines of
$\approx 500$ H~II regions in the outskirts of M81 between long-slit
spectroscopy obtained by \citet{lin2003} and the images obtained by Condor.
(The energy fluxes of the emission lines were summed because the spectral
resolution of the observations obtained by \citealp{lin2003} is insufficient to
resolve \ha\ from [N~II] 658.4 nm.)  Results of the comparison are shown in
Figure 3.  It is evident from Figure 3 that the photometric calibrations of the
\ha\ and [N~II] images in terms of bolometric energy fluxes provide an
excellent match to the spectroscopic observations, thereby demonstrating the
accuracy of the calibrations.

\begin{figure}[ht!]
\centering
\subfloat{
  \includegraphics[width=0.4\linewidth, angle=0]{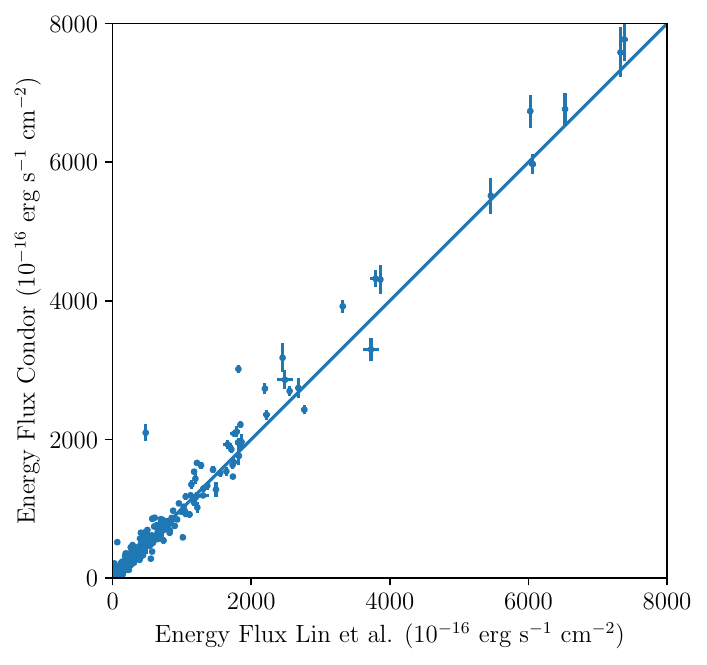}
}
\caption{Comparison of measurements of bolometric energy fluxes of \ha\ plus
[N~II] 658.4 nm emission lines of $\approx 500$ H~II regions in outskirts of
M81 between long-slit spectroscopy obtained by \citet{lin2003} (horizontal
axis) and images obtained by Condor (vertical axis).  Line segment shows equal
energy fluxes.}
\end{figure}

\section{Continuum Subtraction}

The most obvious and dramatic characteristic of the mosaic images obtained
through {\em both} the luminance filter of Figure 1 and the \ha\ filter of
Figure 2 is the intricate web of faint, diffuse, continuum produced by
starlight scattered from Galactic cirrus.  The surface brightness of the cirrus
measured through the luminance filters spans the range $\approx 0.01$ $\mu$Jy
arcsec$^{-2}$ ($\approx 28.9$ mag arcsec$^{-2}$) for the faintest detectable
features through $\approx 0.4$ $\mu$Jy arcsec$^{-2}$ ($\approx 24.9$ mag
arcsec$^{-2}$) for the brightest features.  All prominent cirrus features
identified in the luminance image can also be identified in the \ha\ image (and
in the other narrow-band images).  The cirrus is, of course, interesting in its
own right, but here our primary objective is to search for ionized gas in the
direction of the M81 Group, and toward this end, it is a nuisance.

Indeed one of our motivations for targeting the M81 Group was to experiment
with possibilities of combining deep imaging observations through both broad-
and narrow-band filters to determine and subtract continuum from narrow-band
images in regions of significant Galactic cirrus.  The cirrus is a
near-universal obstacle that all low-surface-brightness imaging
observations---broad- or narrow-band imaging observations of extragalactic or
Galactic objects---must contend with, and it is often the most significant
source of systematic uncertainty associated with deep imaging observations.
Hence it is clearly of interest to find a way to account for the cirrus.

All of the mosaic images are expressed on a common coordinate grid and are
characterized by similar (although not identical) point-spread functions.  And
to at least a first approximation, any line emission is insignificant in
comparison to continuum over the very large bandpass of the luminance filter,
so the luminance image roughly traces continuum, whereas the narrow-band images
trace line emission plus continuum.  Hence we expect that subtracting the
luminance image from the narrow-band images should leave more or less only line
emission in the difference images.

But this is true only to the extent that the continuum is achromatic.  Because
the luminance and narrow-band images are nominally calibrated in AB magnitudes
\citep{oke1983}, ``achromatic'' in this context means of flat specific energy
flux per unit frequency interval $f_\nu$.  This is clearly {\em not} the case
for stars, so we should not expect this procedure to account for stars
perfectly.  And depending on the color of the cirrus and the wavelength
centroid of some narrow-band filter with respect to the luminance filter, it
might be necessary to scale the luminance image by some factor of order unity
to account for the cirrus properly.

With this in mind, we subtracted continuum from each narrow-band image
according to the following procedure:  First, we subtracted the luminance image
from the narrow-band image to form the difference image.  Next, we masked
regions of the difference image around stars contained in the Gaia DR3 catalog
\citep{gai2017, gai2018, gai2021, gai2022} at an isophotal limit, replacing the
values of the masked pixels with the value of the median of nearby pixels.
(This procedure leaves the imprints of faint stars below the limit of Gaia,
which we did not attempt to remove in order to assure that we did not
inadvertently remove bona fide sources of line emission.)  Finally, we repeated
this procedure after scaling the luminance image by a factor of order unity in
order to best account for and remove the cirrus.  This last step is clearly
subjective, but we found through experimentation that it is possible to
determine a scale that accounts for and removes essentially all of the cirrus.

The resulting mosaic difference images obtained through the [O~III], \ha\,
[N~II], and [S~II] filters are shown in Figures 4 through 7, again displayed
block averaged by $32 \times 32$ pixels.  All mosaic difference images (through
the He~II, [O~III], He~I, \ha, [N~II], and [S~II] filters) are available for
download on the Condor web site, as described in § 8 below.

Comparing the mosaic difference images of Figures 4 through 7 with the mosaic
image obtained through the luminance filter shown in Figure 1, it is evident
that the continuum subtraction procedure performed exceptionally well.  The
mosaic difference images show little evidence of residual continuum except
possibly for a very few isolated features.  We conclude that even very
substantial continuum contributed by Galactic cirrus over a very large field of
view can be subtracted to high accuracy given a sufficiently high quality
luminance image to characterize the cirrus; accounting for color variations of
the cirrus (by using images obtained through multiple broad-band filters rather
than only the luminance filter) would likely do better still.

\begin{figure}
\centering
\subfloat{
  \includegraphics[width=0.57\linewidth, angle=0]{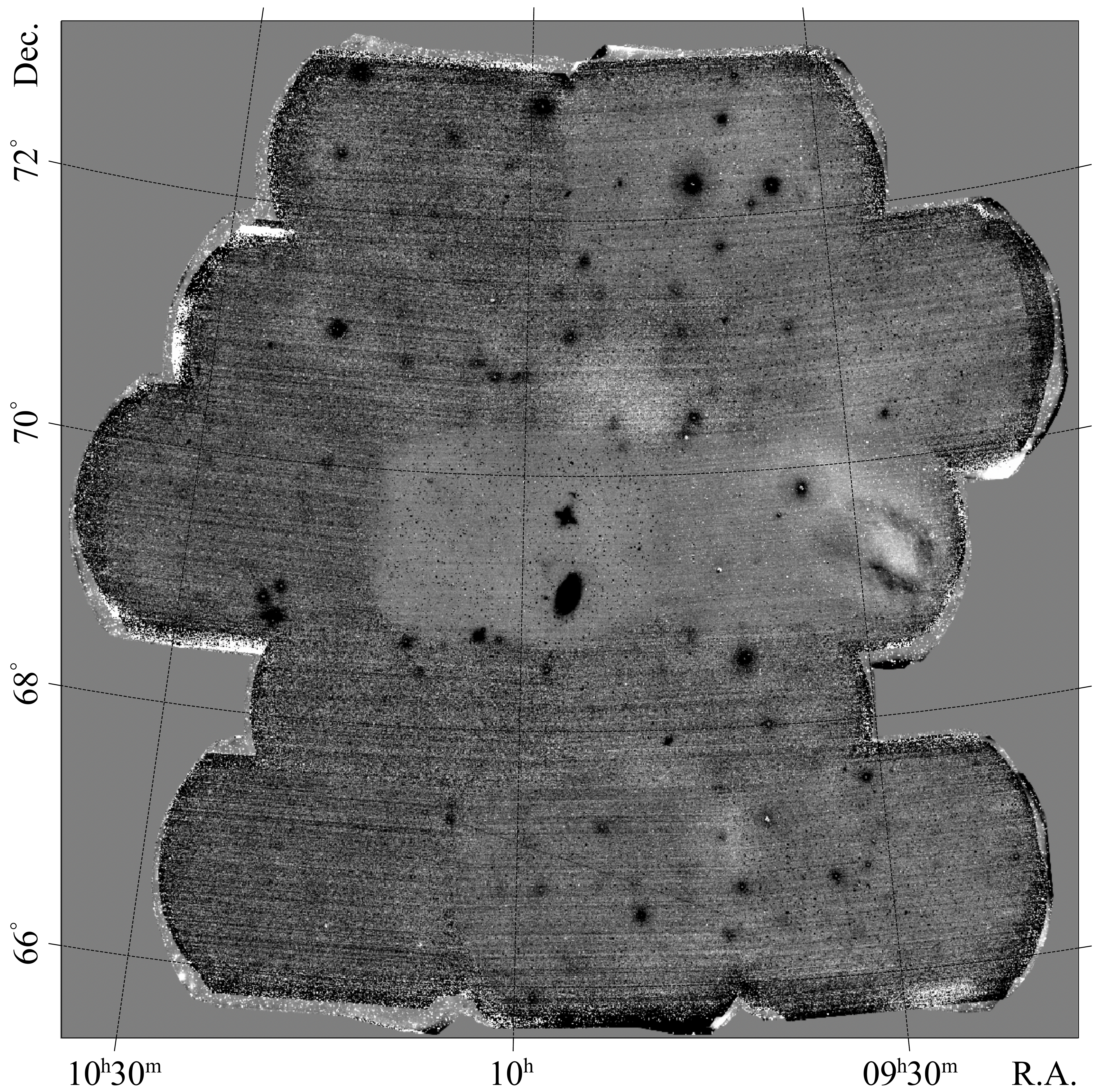}
}
\caption{Mosaic difference image of M81 Group obtained through [O~III] filter
formed by subtracting luminance image from [O~III] image and masking known
stars.  Image is displayed block averaged by $32 \times 32$ pixels.}
\end{figure}

\begin{figure}
\centering
\subfloat{
  \includegraphics[width=0.57\linewidth, angle=0]{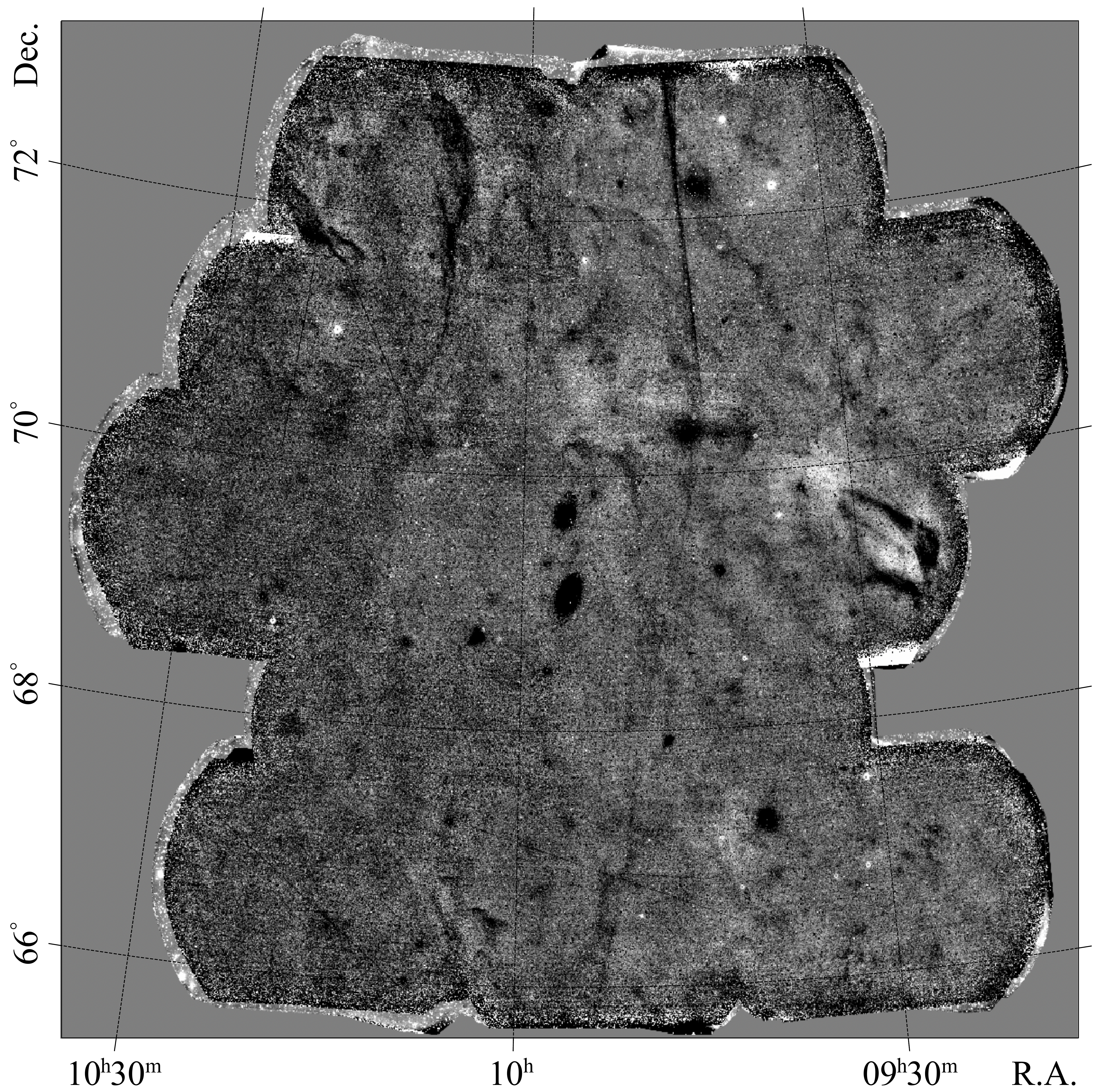}
}
\caption{Mosaic difference image of M81 Group obtained through \ha\ filter
formed by subtracting luminance image from \ha\ image and masking known stars.
Image is displayed block averaged by $32 \times 32$ pixels.}
\end{figure}

\begin{figure}
\centering
\subfloat{
  \includegraphics[width=0.57\linewidth, angle=0]{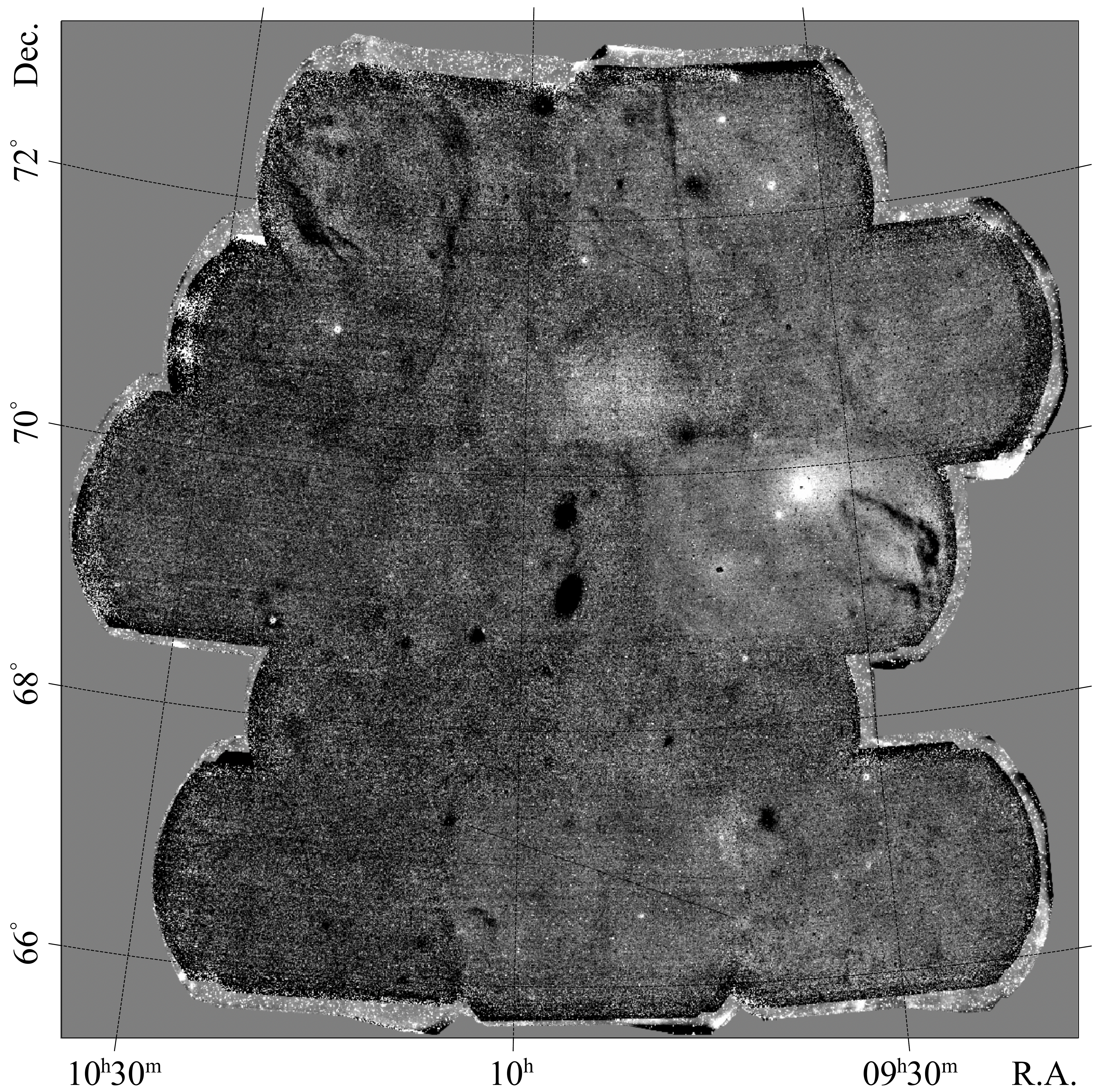}
}
\caption{Mosaic difference image of M81 Group obtained through [N~II] filter
formed by subtracting luminance image from [N~II] image and masking known
stars.  Image is displayed block averaged by $32 \times 32$ pixels.}
\end{figure}

\begin{figure}
\centering
\subfloat{
  \includegraphics[width=0.57\linewidth, angle=0]{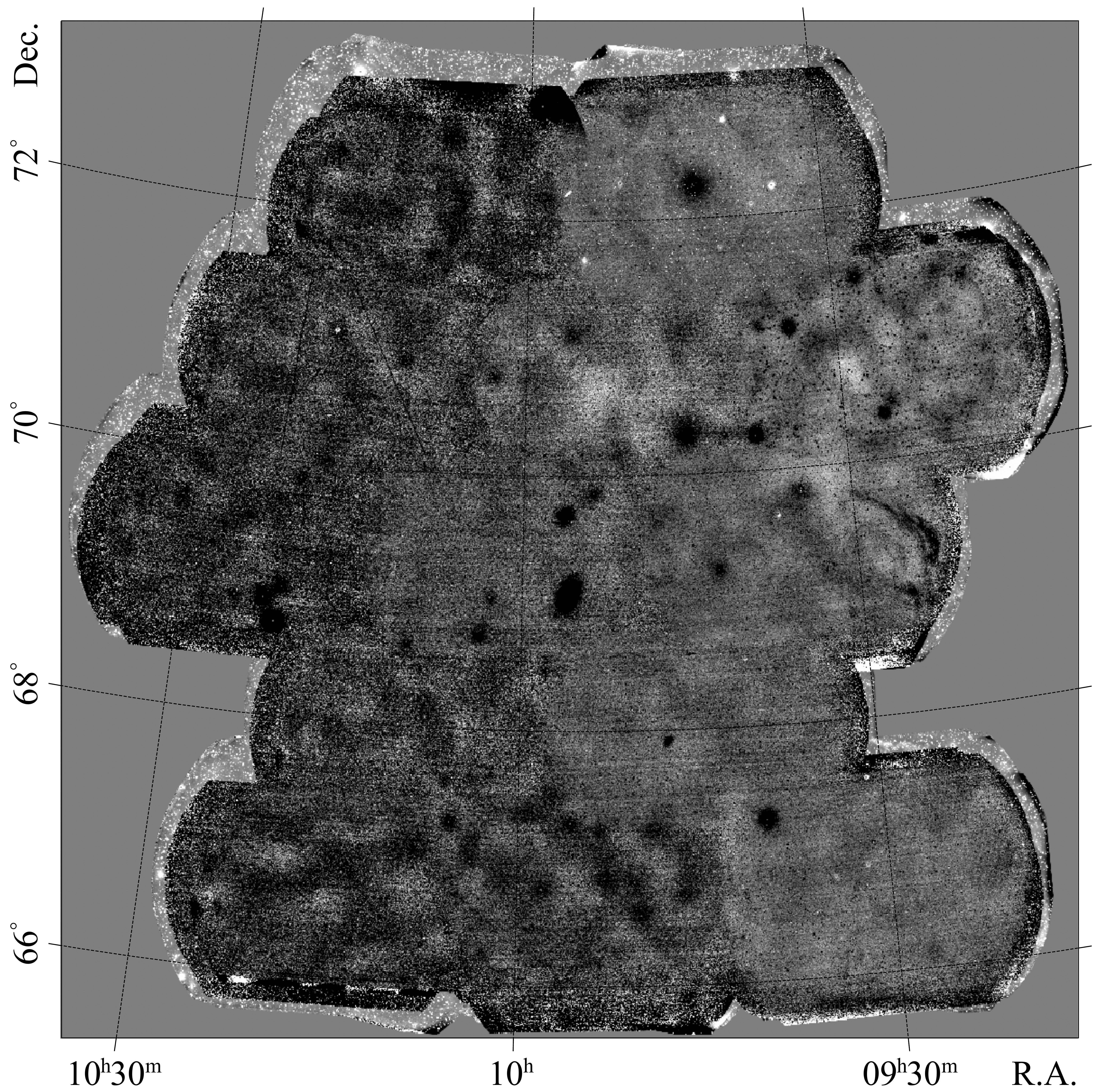}
}
\caption{Mosaic difference image of M81 Group obtained through [S~II] filter
formed by subtracting luminance image from [S~II] image and masking known
stars.  Image is displayed block averaged by $32 \times 32$ pixels.}
\end{figure}

\section{An Overview of Ionized Gas in the Direction of the M81 Group}

The mosaic difference images of Figures 4 through 7 exhibit extensive extended
structures of ionized gas in the direction of the M81 Group.  Emission from
these structures is particularly prominent in \ha\ and [N~II], although
emission from some portions of some of these structures is also evident in
[O~III] and [S~II].  In general, the emission exhibits a complex patchy and
filamentary nature, and it is not at all obvious how to distinguish emission
associated with the M81 Group from emission associated with the Galaxy.  (And
neither is is obvious how spectroscopic observations would resolve this
ambiguity, given the low recession velocity of the group.) Here we present a
brief overview of ionized gas in the direction of the M81 Group, describing
some of the more prominent features visible in the images and trying as best we
can to group features that appear to be related based on location and
morphology.

\newpage

\subsection{Schematic Depiction}

A schematic depiction of some of the extended structures of ionized gas in the
direction of the M81 Group is shown in Figure 8 overlaid upon the \ha\ mosaic
difference image.  Specifically, Figure 8 attempts to depict (1) known galaxies
of the M81 Group (shown as shaded light green circles or ovals), (2) clouds of
gas (shown as shaded light blue circles, ovals, or irregular shapes), (3)
filamentary (or essentially one-dimensional) structures (shown as pink curves
or shaded pink shapes), and (4) apparent or possible bubbles or shells (shown
as open yellow circles) and to exclude known stars and known galaxies not of
the M81 Group.  The features depicted in Figure 8 were drawn by hand,
attempting as best as possible to group objects that appear to be related.
Figure 8 designates some, but not all, of the depicted features or groups of
features with a letter designation.  Figure 8 is not intended to definitively
or exhaustively catalog the ionized gas in the direction of the M81 Group but
rather is intended only to introduce the shapes, sizes, and locations of some
of the extended structures of ionized gas in the direction of the M81 in order
that these structures can be referred to subsequently.  In the following
sections, we call attention to some of the particularly striking and remarkable
features shown in Figure 8; we will present results of the analysis and
interpretation of these observations elsewhere.

\begin{figure}
\centering
\subfloat{
  \includegraphics[width=1.00\linewidth, angle=0]{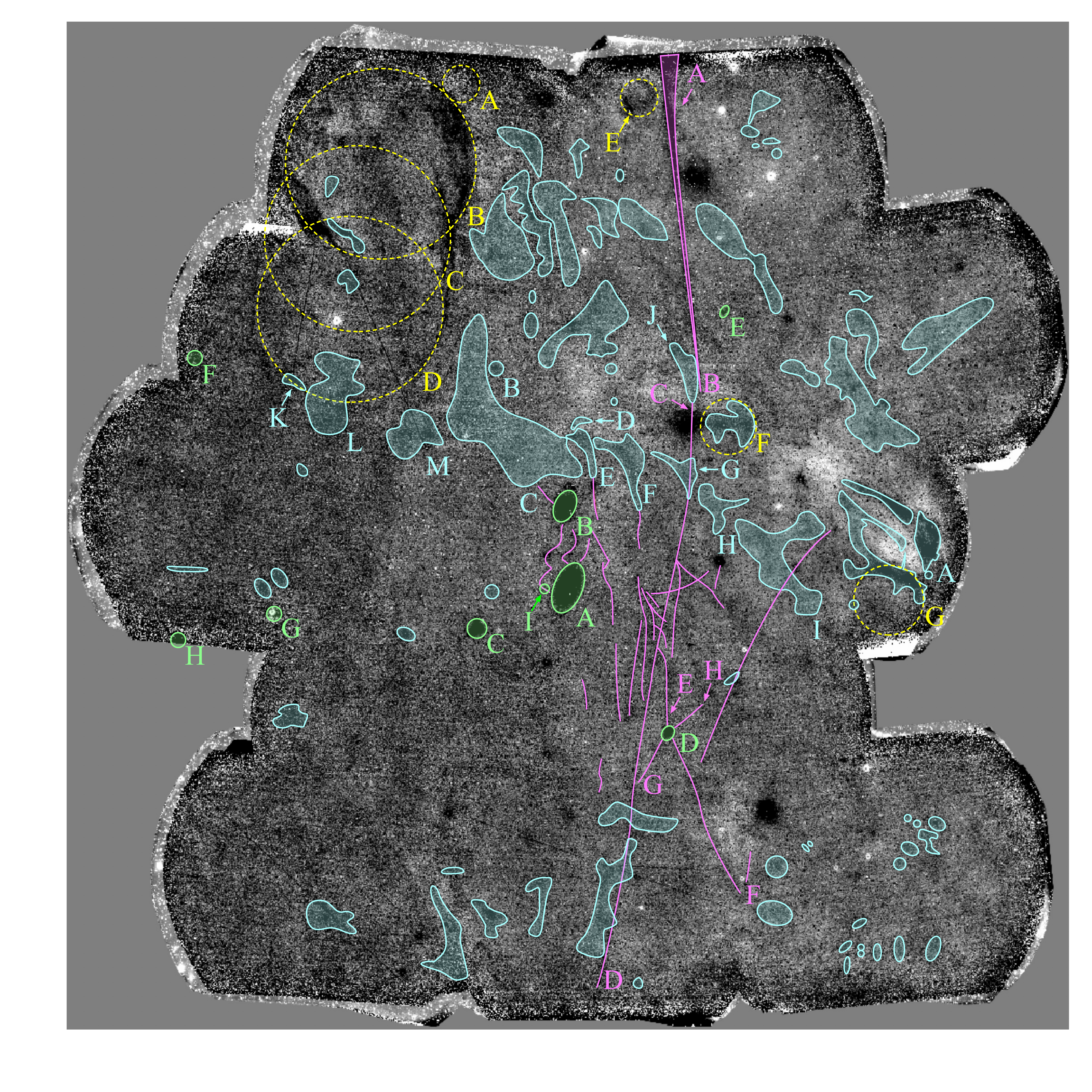}
}
\caption{Schematic depiction of ionized gas in direction of M81 Group.  Known
galaxies of M81 Group are shown as shaded light green circles or ovals), clouds
of gas are shown as shaded light blue circles, ovals, or irregular shapes,
filamentary (or essentially one-dimensional) structures are shown as pink
curves or shaded pink shapes, and apparent or possible bubbles or shells are
shown as open yellow circles.  Known stars and known galaxies not of M81 Group
are not indicated.  Letter designations are positioned to lower right of
corresponding features unless otherwise noted by arrow.  Figure is not intended
to definitively or exhaustively catalog ionized gas in direction of M81 Group
but rather is intended only to introduce shapes, sizes, and locations of some
of extended structures of ionized gas in direction of M81 in order that these
structures can be referred to subsequently.}
\end{figure}

\subsection{Galaxies of the M81 Group}

Eight known galaxies of the M81 Group are visible in one or more of the mosaic
difference images.  Details of these galaxies are presented in Table 3, which
for each galaxy lists the galaxy name, designation in Figure 8, and ICRS J2000
coordinates.  There are other galaxies of the M81 Group that are not visible in
the mosaic difference images, and there are other galaxies not of the M81 Group
that are visible in the mosaic difference images.

\begin{table}[ht]
\centering
\hspace{-0.80in}
\begin{tabular}{p{1.5in}ccc}
\multicolumn{4}{c}{{\bf Table 3:}  Details of Galaxies Visible in Images} \\
\hline
\hline
\multicolumn{1}{c}{} & & \multicolumn{2}{c}{J2000} \\
\cline{3-4}
\multicolumn{1}{c}{Galaxy} & Designation & R.A. & Dec \\
\hline
M81 \dotfill               & A & 09:55:33.2 & $+$69:03:55.1 \\
M82 \dotfill               & B & 09:55:52.4 & $+$69:40:46.9 \\
NGC 3077 \dotfill          & C & 10:03:19.1 & $+$68:44:01.6 \\
NGC 2976 \dotfill          & D & 09:47:15.5 & $+$67:54:59.0 \\
UGC 5139 \dotfill          & E & 09:40:32.3 & $+$71:10:56.0 \\
UGC 5692 \dotfill          & F & 10:30:35.0 & $+$70:37:07.2 \\
HIJASS J1021+68 \dotfill   & G & 10:21:00.2 & $+$68:42:00.0 \\
IC 2574 \dotfill           & H & 10:28:23.6 & $+$68:24:43.4 \\
UGC 5536 \dotfill          & I & 09:57:32.9 & $+$69:02:50.7 \\
\hline
\end{tabular}
\end{table}

\subsection{Ursa Major Arc}

The mosaic difference images show a remarkable filament known as the ``Ursa
Major Arc'' \citep{mcc2001, bra2020a} that runs roughly N-S across the entire
field of view.  The filament is visible in the \ha\ and [N~II] mosaic
difference images.  The filament is designated in Figure 8 as comprised of two
pieces:  a northern piece that runs between filament designations A and B, and
a southern piece that runs between filament designations C and D.  The
character of the filament changes across the \ha\ and [N~II] images in the
sense that it appears thicker toward the N and thinner toward the S, and the
filament appears to exhibit a discontinuity coincident with cloud J.  The Ursa
Major Arc is known to stretch $\approx 30$ deg on the sky \citep{bra2020a} and
has been interpreted variously as a trail of ionized gas produced by an unseen
ionizing source \citep{mcc2001} and as an interstellar shock \citep{bra2020a}.
It is notable that the Ursa Major Arc passes within only $\approx 0.75$ deg of
the center of M81.

\subsection{Additional Filaments}

The mosaic difference images show a remarkable network of criss-croseds
filaments, some of which intersect and overlap the Ursa Major Arc.  The
filaments are visible in the \ha\ mosaic difference image, and some of the
filaments are visible in the [N~II] mosaic difference image.  The epicenter of
the network of filaments appears to be located $\approx 0.9$ deg SW of M81,
between M81 and NGC 2976.  The filaments are too numerous to cleanly designate
individually in Figure 8, so only some of the filaments are designated.  At
least three of the filaments appear to originate or terminate on NGC 2976: one
at roughly 12 o'clock (designated filament E), one at roughly 5 o'clock
(filament F), and one at roughly 7 o'clock (filament G).  Filaments E and G
(as well as some of the undesignated filaments) appear to intersect and overlap
the Ursa Major Arc.

\subsection{Giant Shell of Ionized Gas}

The mosaic difference images show the ``giant shell of ionized gas'' discovered
by \citet{lok2022} located $\approx 0.6$ deg to the NW of M82.  The giant shell
is visible in the \ha\ and [N~II] mosaic difference images.  These mosaic
difference images reveal several new aspects of the giant shell:
\begin{enumerate}

\item The giant shell is clearly comprised of at least two pieces; an eastern
piece designated in Figure 8 as cloud E, and a western piece designated as
cloud F.  The western-most edge of cloud F is located within $\approx 0.35$ deg
of the Ursa Major Arc.

\item Both pieces of the giant shell exhibit similar morphologies, with a shape
that resembles a ``comma.''  Further, both pieces appear to extend toward the
S in thin filaments.  These filaments form part of the network of
criss-crossed filaments that intersect and overlap the Ursa Major Arc described
above.

\item There is another cloud designated as cloud G located immediately to the W
of cloud F that exhibits a morphology similar to those of clouds E and F, with
a shape that also resembles a comma.  Cloud G is essentially coincident with
the Ursa Major Arc.

\item In fact, clouds E, F, and G appear to form part of a larger chain of
clouds that run roughly NE to SW that includes clouds designated as clouds K,
L, M, B, C, and D to the E and clouds H and I to the W.  It is, of course,
uncertain whether these clouds are physically related or even at the same
distance.

\end{enumerate}

\subsection{Vicinity of M81 and M82}

The mosaic difference images show various features in the vicinity of M81 and
M82, including the well-known ``\ha\ cap'' of M82 \citep{dev1999, leh1999}
located $\approx 9$ arcmin NW of the center of M82, the ``\ha\ ridge''
discovered by \citet{lok2022} located roughly midway between the outer extent
of M82 and the \ha\ cap, and the ``\ha-emitting filament'' the ``\ha-emitting
clump'' noted previously by \citet{pas2021}.  The features are visible in the
\ha\ mosaic difference image, and some of the features are visible in the
[O~III], [N~II], and [S~II] mosaic difference images.  Some of the features
cannot be resolved from M82 in the mosaic difference images shown in Figures 4
through 7 but are easily visible on the images displayed block averaged on a
finer scale.

\subsection{Rattlesnake Head Nebula}

The mosaic difference images show an apparent emission nebula located near the
western edge of the field of view, due almost exactly W of the midpoint between
M81 and M82.  The nebula is visible in all four of the mosaic difference
images.  The nebula appears to be comprised of several distinct pieces, which
are collectively designated in Figure 8 as cloud A.  The shape of the nebula
reminds us of the head of a rattlesnake, and we dub the nebula the
``Rattlesnake Head Nebula.''

\subsection{Apparent or Possible Bubbles or Shells}

The mosaic difference images show several apparent or possible bubbles or
shells.  The bubbles or shells are visible in the \ha\ mosaic difference image,
and some of the bubbles or shells are visible in the [N~II] mosaic difference
image.  A particularly striking apparent bubble or shell designated in Figure 8
as bubble B comprises two near halves of nearly-circular feature in the
upper-left corner of the image, with a center due NE of M81 and M82 and a
radius of $\approx 0.75$ deg.  A second particularly striking apparent bubble
or shell designated as bubble D is located due S of bubble B, also with a
radius of $\approx 0.75$ deg.  A possible piece of a bubble or shell designated
as bubble C is located between bubbles B and D.  A possible bubble or shell
designated as bubble F is located near the discontinuity of the Ursa Major Arc
coincident with cloud J, with a radius of $\approx 0.2$ deg.  And a possible
bubble or shell designated as bubble G is located just S of the Rattlesnake
Head Nebula.  Other possible bubbles or shells designated as A and E are
located N of M81 and M82.

\subsection{Other Emission Features}

The mosaic difference images show a large number of other emission features, of
a variety of sizes, shapes, and brightnesses.  Although we do not individually
enumerate all of these other features here, they can be picked out in the
mosaic difference images of Figures 4 through 7 or in the images available for
download on the Condor web site, as described in \S\ 8 below.  The various
emission features visible in the mosaic difference images could be targets of
future spectroscopic observations with large ground-based telescopes, although
as noted above, the very low recession velocity of the M81 Group will make it
difficult to distinguish features of Galactic origin from features at the
distance of the group.

\section{Summary}

We used the Condor Array Telescope to obtain deep imaging observations through
the luminance broad-band and He~II 468.6 nm, [O~III] 500.7 nm, He~I 587.5 nm,
\ha, [N~II] 658.4 nm, and [S~II] 671.6 nm narrow-band filters of an extended
region comprising 13 ``Condor fields'' spanning $\approx 8 \times 8$ deg$^2$ on
the sky centered near M81 and M82.  Here we describe the acquisition and
processing of these observations, which together constitute unique very deep
imaging observations of a large portion of the M81 Group through a complement
of broad- and narrow-band filters.  The images obtained through the broad-band
filter and the narrow-band filters are characterized by an intricate web of
faint, diffuse, continuum produced by starlight scattered from Galactic cirrus,
and all prominent cirrus features identified in the broad-band image can also
be identified in the narrow-band images.  We subtracted the luminance image
from the narrow-band images to leave more or less only line emission in the
difference images, and we masked regions of the resulting images around stars
at an isophotal limit.

The mosaic difference images exhibit extensive extended structures of ionized
gas in the direction of the M81 Group, from known galaxies of the M81 Group,
clouds of gas, filamentary structures, and apparent or possible bubbles or
shells.  Specifically, the difference images show a remarkable filament known
as the ``Ursa Major Arc;'' a remarkable network of criss-crossed filaments
between M81 and NGC 2976, some of which intersect and overlap the Ursa Major
Arc; details of a ``giant shell of ionized gas;'' and a large number of other
emission features.

\section{Data Availability}

All raw Condor data are available following an 18-month proprietary period.
All raw and processed data described here, including the coadded mosaic images
of and the mosaic difference images, are available on the Condor web site
https://condorarraytelescope.org/data\_access/ or by contacting the
corresponding author.

\begin{acknowledgments}
This material is based upon work supported by the National Science Foundation
under Grants 1910001, 2107954, 2108234, 2407763, and 2407764.  We gratefully
acknowledge the staff of Dark Sky New Mexico, including Diana Hensley and
Michael Hensley for their superb logistical and technical support.  The authors
thank the anonymous referee for very valuable comments.
\end{acknowledgments}

\software{
  astroalign \citep{ber2020},
  astropy \citep{2013A&A...558A..33A,2018AJ....156..123A},
  django \citep{django},
  Docker \citep{mer2014a},
  DrizzlePac \citep{gon2012},
  NoiseChisel \citep{akh2015, akh2019},
  numba \citep{lam2015},
  numpy \citep{har2020},
  photutils \citep{bra2020},
  scipy \citep{vir2020},
  SExtractor \citep{1996A&AS..117..393B}
}

\bibliography{manuscript}{}
\bibliographystyle{aasjournal}

\end{document}